\documentclass[%
 aip,
 amsmath,amssymb,
 reprint,%
]{revtex4-1}

\usepackage{graphicx}
\usepackage{dcolumn}
\usepackage{bm}

\usepackage[utf8]{inputenc}
\usepackage[T1]{fontenc}
\usepackage{mathptmx}
\usepackage{etoolbox}

\makeatletter
\def\@email#1#2{%
 \endgroup
 \patchcmd{\titleblock@produce}
  {\frontmatter@RRAPformat}
  {\frontmatter@RRAPformat{\produce@RRAP{*#1\href{mailto:#2}{#2}}}\frontmatter@RRAPformat}
  {}{}
}%
\makeatother
\begin{document}

\preprint{AIP/123-QED}

\title{Feasibility study of two-stage direct laser acceleration of electrons}
\author{E. Starodubtseva}
 \email{starodubtceva.em19@physics.msu.ru}
\affiliation{Lomonosov Moscow State University, Faculty of Physics, 119991, Moscow, Russia}
\affiliation{Institute for Nuclear Research of Russian Academy of Sciences, 117312, Moscow, Russia}

\author{S. Solntsev}
 \affiliation{Lomonosov Moscow State University, Faculty of Physics, 119991, Moscow, Russia}
\affiliation{Lomonosov Moscow State University, Sarov Branch, 607328, Sarov, Russia}

\author{I. Tsymbalov}
\affiliation{Lomonosov Moscow State University, Faculty of Physics, 119991, Moscow, Russia}
\affiliation{Institute for Nuclear Research of Russian Academy of Sciences, 117312, Moscow, Russia}

\author{A. Savel'ev}
\affiliation{Lomonosov Moscow State University, Faculty of Physics, 119991, Moscow, Russia}
\affiliation{Lomonosov Moscow State University, Sarov Branch, 607328, Sarov, Russia}
\affiliation{Lebedev Physical Institute of Russian Academy of Sciences, 119192, Moscow, Russia}

\date{\today}

\begin{abstract}
We present the first numerical demonstration of two-stage direct laser acceleration (DLA) of electrons. The result was obtained using quasi-3D particle-in-cell (PIC) simulations. The first stage (\(a_0 = 6\), \(\tau = 200~\mathrm{fs}\), \(w = 4\lambda\), \(\lambda = 0.8~\mu\mathrm{m}\)) utilized a nitrogen gas target with an atomic density of \(n_a = 0.008n_{cr}\). This stage was designed to accelerate low-energy electrons, achieving a maximum energy of approximately 94~MeV with a two-temperature energy spectrum (\(T_1 = 7.0\) MeV and \(T_2 = 16.0\) MeV) and a charge of 6.04~nC (\(\mathcal{E} > 5\) MeV). The resulting electron bunch was then injected into the second stage, where hydrogen at a lower density (\(n_a = 0.01n_{cr}\)) was used, while the laser parameters were the same as in the first stage except that the waist was increased to \(w = 8\lambda\). After the second stage, the electrons reached energies up to 225~MeV, with a two-temperature energy spectrum (\(T_1 = 18.4\) MeV and \(T_2 = 88.3\) MeV) and a charge of 4.55~nC (\(\mathcal{E} > 5\) MeV). The self-injected electrons in the second stage reached a maximum energy of 123~MeV with a charge of 1.14~nC (\(\mathcal{E} > 5\) MeV). The simulation results are compared with a theoretical DLA model, which describes the electron motion in the combined fields of the laser pulse and the quasistatic plasma channel. The model accounts for the transverse size of the channel by incorporating a Gaussian attenuation of the channel fields and off-axis injection. The theoretical model correctly describes the observed acceleration dynamics for both low-energy and pre-accelerated electrons, confirming its validity for designing multi-stage DLA schemes. Thus, the implementation of multi-stage acceleration by the DLA mechanism is successfully demonstrated for the first time in a numerical experiment.
\end{abstract}

\maketitle

\section{\label{sec:level1}Introduction}

Laser-plasma accelerators are among the most rapidly developing fields in accelerator physics \cite{esarey2009physics}. Owing to their ability to sustain accelerating gradients three to four orders of magnitude higher than those of conventional RF accelerators \cite{hooker2013developments}, these systems offer the possibility of creating sources of relativistic electron beams for a wide range of applications, including nuclear photonics, medical diagnostics and therapy, X-ray imaging, as well as compact sources of coherent radiation for next-generation free-electron lasers \cite{nedorezov2021nuclear, albert2016applications, corde2013femtosecond, galletti2024prospects}.

Among the mechanisms of laser-plasma acceleration, two are most extensively studied: laser wakefield acceleration (LWFA), in which electrons are accelerated by the longitudinal electric field of a plasma wave excited by an laser pulse \cite{tajima1979laser, pukhov2002laser, faure2004laser, esarey2009physics, gonsalves2019petawatt}, and direct laser acceleration (DLA), in which electrons gain energy directly from the laser field while propagating in a plasma channel \cite{pukhov1999particle, arefiev2016beyond, arefiev2014enhancement, rosmej2020high, hussein2021towards, tsymbalov2019well, tsymbalov2020efficient}.

The maximum energy gain attainable within a single stage of laser-plasma electron accelerator is subject to fundamental constraints. In the case of LWFA, the principal limitations are electron dephasing, which arises from the subluminal phase velocity of the plasma wave and results in the accelerated electrons outrunning the accelerating phase of the wakefield, whereupon they begin to decelerate \cite{esarey2009physics, lu2007generating, tajima1979laser}, and pump depletion, whereby the laser pulse progressively transfers its energy to the wakefield until it is exhausted, thereby terminating the acceleration process \cite{esarey2009physics, lu2007generating, esarey2004nonlinear}.

The acceleration process in DLA mechanism is bounded by the finite longitudinal extent of the plasma channel, which persists only for as long as the laser pulse maintains sufficient intensity to sustain it; upon channel dissipation, the acceleration ceases \cite{pukhov1999particle, gahn1999multi}. Furthermore, efficient energy gain in DLA necessitates the fulfilment of a resonance condition between the betatron oscillation frequency and the Doppler-shifted laser frequency \cite{pukhov1999particle, tsakiris2000laser, arefiev2016beyond}. Once this condition is satisfied, the electron undergoes continuous acceleration; however, upon reaching the maximum energy dictated by the channel parameters, the electron enters a deceleration regime and begins to loose energy to the laser field. Consequently, even in an idealized infinitely long channel, the electron energy cannot be increased indefinitely \cite{tsakiris2000laser, starodubtseva2023low}.

These limitations are intrinsic to the single-stage configuration and cannot be circumvented solely by increasing the laser power or extending the plasma length \cite{leemans2009laser, esarey2009physics}. Multi-stage acceleration is a natural solution to the limitations outlined above: by partitioning the acceleration process into several independently driven modules, each stage can operate at its optimal length, and the electron beam can be transferred from one stage to the next prior to the onset of dephasing, depletion, channel dissipation, or deceleration \cite{leemans2009laser, schroeder2010physics}. This concept has been proposed and extensively studied for the LWFA mechanism. Several numerical studies have explored multi-stage LWFA schemes and their coupling requirements \cite{reitsma2002simulation, haq2025multi, maity2025coupling}. Experimentally, multi-stage acceleration has also been realized. The first demonstration of a staged all-optical LWFA with two separate laser pulses was reported in 2005 \cite{kaganovich2005first}, where electrons from a 2 TW laser pulse were injected into a second stage driven by a 10 TW laser pulse and accelerated beyond 20 MeV. Two-stage coupling with separate laser pulses (12 TW and 28 TW) using a plasma mirror and active plasma lens achieved an additional 100 MeV gain during the second stage, reaching total energies above 220 MeV \cite{steinke2016multistage}. A two-stage scheme with separate laser pulses (17 TW and 12 TW) and a $\sim1$ m gap between stages was reported in \cite{jin2019coupling}, where 10 MeV bunches were injected into the booster and accelerated to 30 MeV. In all these experiments, stages were coaxial: the laser pulse of the second stage propagated along the same axis as the electron beam from the first stage.

Motivated by this experimentally established configuration, the same coaxial geometry is adopted in the present work. However, whereas the experiments above addressed multi-stage acceleration within the LWFA mechanism, here we consider multi-stage acceleration based on the DLA mechanism, which offers a number of advantages that are particularly attractive for multi-stage schemes. The record values of charge per unit laser energy for DLA are substantially higher than those for LWFA, reaching the level of several nanocoulombs per joule across a wide range of laser parameters and target configurations \cite{rosmej2020high, rosmej2019interaction, tsymbalov2020efficient, tsymbalov2019well, shaw2021high}. The energy conversion efficiency from laser to electrons follows the same trend: in DLA, the conversion efficiency reaches 23\% for electrons above 2 MeV \cite{rosmej2020high}, whereas typical LWFA experiments exhibit conversion efficiencies of only a few percent \cite{faure2004laser, mangles2004monoenergetic, geddes2004high, ke2021near, gilljohann2020physics, foerster2022stable}. These properties make DLA particularly well suited for multi-stage acceleration schemes, where high charge and efficient energy transfer are of primary importance.

We have previously proposed a multi-stage DLA scheme based on an analytical model of the DLA mechanism \cite{starodubtseva2023cascade}, showing that this DLA is particularly well suited for multi-stage acceleration due to its specific feature: depending on the plasma channel parameters, both low-energy and pre-accelerated electrons can be efficiently accelerated by different regimes of DLA. We consider that this property will enable more efficient multi-stage DLA compared to the LWFA-based staging schemes.

Multi-stage electron acceleration is of particular relevance to emerging multi-channel mega-science facilities such as XCELS in Russia \cite{khazanov2023exawatt} and ELI Beamlines in Europe \cite{tesileanu2016laser}, which are explicitly designed to deliver multiple synchronized laser pulses for staged acceleration. This stimulates interest in the study of multi-stage electron acceleration. For LWFA, this problem has been studied only to a limited extent, while for DLA it has not been studied at all.

In this work, we present the first numerical demonstration of two-stage direct laser acceleration (DLA) of electrons in quasi-3D particle-in-cell (PIC) simulations. The first stage (\(a_0 = 6\), \(\tau = 200\) fs, \(w = 4\lambda\), \(n_e = 0.056n_{cr}\), \(\lambda = 0.8~\mu\mathrm{m}\), peak power \(\sim 9\) TW, pulse energy \(1.8\) J) was designed to accelerate low-energy electrons, achieving a maximum energy of approximately 94 MeV with a two-temperature energy spectrum (\(T_1 = 7.0\) MeV and \(T_2 = 16.0\) MeV) and a charge of \(6.04\) nC (for electrons with energies above 5 MeV). The resulting electron bunch was then injected into the second stage (\(a_0 = 6\), \(\tau = 200\) fs, \(w = 8\lambda\), \(n_e = 0.01n_{cr}\), \(\lambda = 0.8~\mu\mathrm{m}\)), where the laser waist was increased and the plasma density was lowered to efficiently accelerate pre-accelerated electrons. After the second stage, the electrons reached energies up to 225 MeV, with a two-temperature energy spectrum (\(T_1 = 18.4\) MeV and \(T_2 = 88.3\) MeV) and a preserved charge of \(4.55\) nC (for electrons with energies above 5 MeV). In contrast, the self-injected electrons in the second stage reached a maximum energy of 123 MeV with a charge of \(1.14\) nC (for electrons with energies above 5 MeV). The simulation results confirm the feasibility of sequential DLA acceleration and demonstrate that the DLA mechanism, despite its high charge and efficiency, can be successfully implemented in a multi-stage configuration, where the total accelerated charge and maximum electron energies are significantly higher than those attainable in a single-stage scheme.

These parameters were chosen with several considerations in mind. First, they are sufficient to form a stable plasma channel that persists over an extended propagation distance, which is a prerequisite for the DLA mechanism \cite{pukhov1999particle, gahn1999multi}. Second, the computational cost of such simulations remains manageable. The acceleration length in DLA scales with laser power as \(L_{\text{acc}} \propto P^{1/3}\), so a transition from the second stage (36~TW) to the 10~PW class increases it by a factor of seven \cite{babjak2024direct}. Assuming a typical DLA divergence of \(\sim\)100~mrad \cite{rosmej2020high}, the transverse bunch size would grow proportionally, increasing the computational cost by approximately two orders of magnitude. Third, lasers with the parameters chosen in this work are representative of systems that can operate at 10~Hz. Such repetition rates would enable the geometry and injection scheme of the present work to be tested experimentally and optimized before scaling to more powerful multi-petawatt facilities such as XCELS \cite{khazanov2023exawatt} and ELI Beamlines \cite{tesileanu2016laser}.

The simulation results obtained in this work were compared with the theoretical DLA model, which was upgraded from our previous study~\cite{starodubtseva2023cascade} to account for the finite transverse size of the electron beam and the plasma channel. This upgrade incorporates a Gaussian attenuation of the channel fields and off-axis injection. The good agreement between the PIC-simulations and the analytical predictions confirms that the upgraded model can be used for estimating the optimal parameters of multi-stage DLA schemes. To the best of our knowledge, the numerical demonstration of two-stage DLA acceleration reported here is the first of its kind. These results establish a foundation for the development of multi-stage accelerators based on the DLA mechanism, which are characterized by high charge and relaxed synchronization requirements in comparison with LWFA-based staging schemes.

\section{First-stage acceleration}

The simulations were performed using the open-source PIC-code Smilei~\cite{derouillat2018smilei} in cylindrical geometry with azimuthal modes decomposition~\cite{lifschitz2009particle}. The use of two azimuthal modes (\(m=0,1\)) is sufficient for the axially symmetric problem under consideration, as was systematically validated in our recent work \cite{nikiforov2026azimuthal}. The grid resolution was \(\lambda/32\) longitudinally and \(\lambda/8\) radially, with a timestep of \(\lambda/48c\) and \(\lambda = 0.8~\mu\mathrm{m}\). The simulation domain was \(250\lambda \times 40\lambda\), with a total simulation time of \(370\lambda/c\). The laser pulse had the normalized vector potential \(a_0 = 6\), duration of \(\tau = 200\) fs (FWHM), and waist of \(w = 4\lambda\) (FWHM), corresponding to a peak power of approximately \(9\) TW and a pulse energy of \(1.8\) J. The choice of a relatively long pulse duration of 200 fs was motivated by the requirements of the DLA mechanism: while shorter pulses are preferable for LWFA, efficient DLA requires the formation of a sufficiently long plasma channel that persists over an extended propagation distance, which is achieved with longer pulses. At the same time, excessively long pulses would lead to an increase in the simulation domain and computational cost, and, more importantly, to a significant increase in the pulse energy at the given \(a_0\), since the energy scales linearly with the pulse duration. A duration of 200 fs was therefore found to provide a reasonable compromise, ensuring both the formation of a stable plasma channel, manageable laser pulse energy and computational requirements. 

In the first stage PIC-simulation nitrogen was used as the target gas. The profile of the atomic nitrogen density was shaped as a Gaussian up-ramp with a plateau at \(n_{a0} = 0.008n_{cr}\): \(n_a = n_{a0}\exp[-((x - 41\lambda)/(15\lambda))^2]\) for \(x < 41\lambda\) and \(n_a = n_{a0}\) for \(x \geq 41\lambda\). At the laser intensity employed in this work, all seven electrons of the nitrogen atom are ionized, so that the electron density reaches \(n_e = 0.056n_{cr}\). The ionization dynamics were described using the ADK model \cite{ammosov1986tunnel}. A systematic scan over the electron density showed that lower densities yielded weaker acceleration and higher densities suppressed channel formation; \(n_e = 0.056n_{cr}\) provided the optimal balance.

\begin{figure*}
\includegraphics[width=1.0\linewidth]{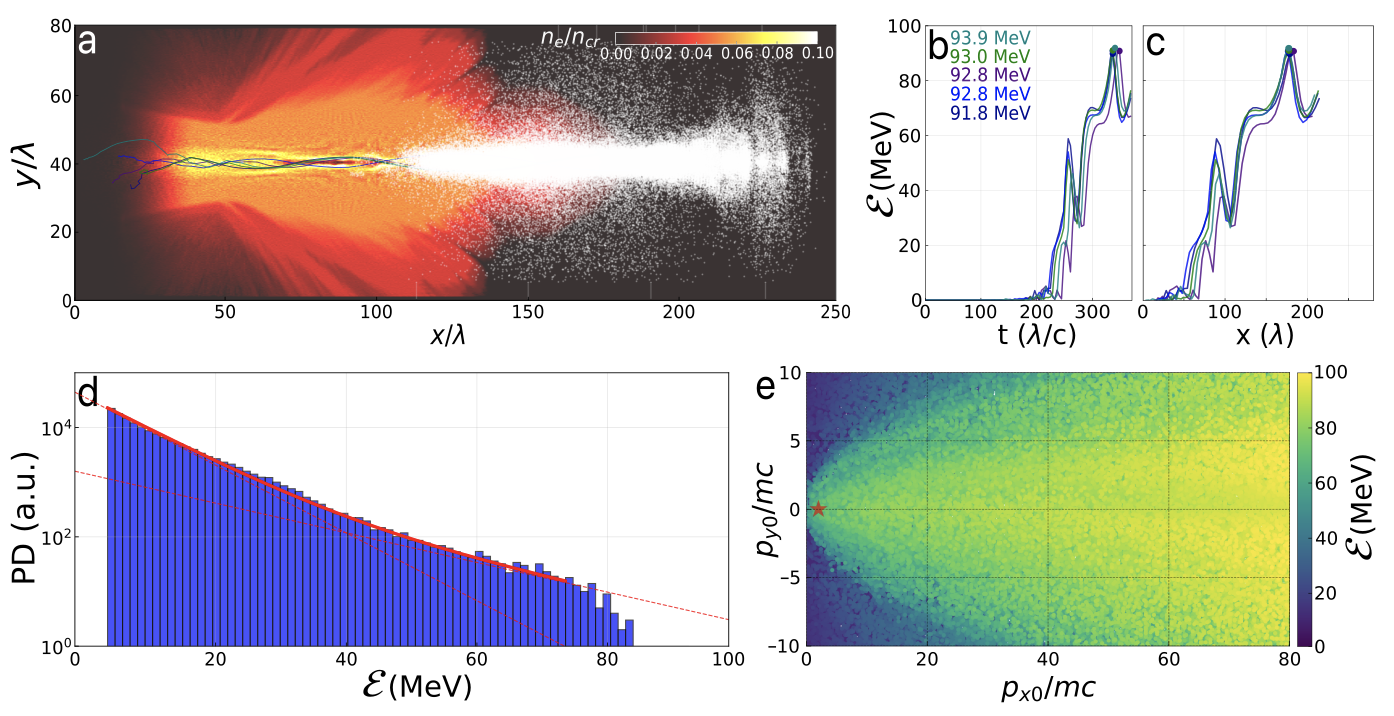}
\centering
\caption{\label{Figure1} First-stage acceleration results. (a) Plasma density snapshot at \(t = 240\lambda/c\), overlaid with trajectories of the five most energetic electrons. The white scatters show the electron bunch with \(p_x/mc > 5\) at the final instant, which is subsequently injected into the second stage. (b) Energy evolution over time for the five most energetic electrons; the maximum energy reached by each electron is indicated in the legend, dots indicate the instants of maximum energy. (c) Energy of the same electrons as a function of longitudinal position; dots indicate the positions of maximum energy. (d) Energy spectrum of accelerated electrons above 5 MeV (blue histogram) and fit \(dN/d\mathcal{E} \propto A_1 \exp(-\mathcal{E}/T_1) + A_2 \exp(-\mathcal{E}/T_2)\) (solid red curve), yielding \(T_1 = 7.0\) MeV and \(T_2 = 16.0\) MeV (dashed red curves). (e) Maximum energy of test electrons as a function of initial longitudinal and transverse momenta. The test electrons were injected at \(x_0 = (25-27)\lambda\), \(r_0 = 0.5\lambda\), \(t_0 = 200\lambda/c\) with small weights to avoid perturbing the plasma. The white dashed curves mark the analytical boundaries of the different acceleration regime. The red star indicates the point \((p_x/mc, p_y/mc) = (3, 0)\) selected for phase portrait analysis (see Sec.\ref{Sec3}). Simulation domain: \(250\lambda \times 40\lambda\), total time \(370\lambda/c\).}
\end{figure*}

The principal results of the first-stage acceleration are summarized in Fig.~\ref{Figure1}. Fig.~\ref{Figure1}a displays a snapshot of the plasma density at \(t = 240\lambda/c\), overlaid with the trajectories of the five most energetic electrons. The white scatters show the electron bunch with \(p_x/mc > 5\) at the final instant, which is subsequently injected into the second stage. The figure reveals a well-defined, stable plasma channel -- a region substantially depleted of electrons along the propagation axis -- surrounded by dense plasma sheaths. Here, the accelerated electrons are self-injected from the channel sheaths (in contrast to the artificially injected electrons considered below). The trajectories of these electrons exhibit pronounced sinusoidal oscillations, which are the betatron oscillations and serve as a distinctive signature of the DLA mechanism. Fig.~\ref{Figure1}b illustrates the temporal evolution of the energy for the five most energetic electrons, demonstrating the acceleration process at approximately \(150\lambda/c\). Fig.~\ref{Figure1}c shows the energy of the same electrons as a function of their longitudinal coordinate demonstrating the acceleration process at approximately \(140\lambda\). The labels in panel Fig.~\ref{Figure1}b indicate the maximum energy reached by each electron, while the markers in panels Fig.~\ref{Figure1}b and Fig.~\ref{Figure1}c highlight the instants and positions at which the maximum energy is achieved for the corresponding electrons. Fig.~\ref{Figure1}d presents the energy spectrum of the accelerated electron beam for energies exceeding 5 MeV. The spectrum, shown as a blue histogram, which is typical of DLA, is well described by the function \(dN/d\mathcal{E} \propto A_1\exp(-\mathcal{E}/T_1)+A_2\exp(-\mathcal{E}/T_2)\) (solid red curve), with an effective temperatures \(T_1 = 7.0\) MeV and \(T_1 = 16.0\) MeV. The maximum energy reaches 94 MeV, while the integrated charge of the beam amounts to 6.04~nC for electrons with energies higher than 5 MeV. Such a two-temperature (bi-exponential) form of the DLA spectrum is consistent with the earlier experimental observations in near-critical-density plasmas \cite{rosmej2020high,rosmej2019interaction}. A theoretical interpretation of this structure was given in Ref.~\cite{veysman2024estimation}, where it was discussed that the DLA electron energy distribution can be approximated by a two-temperature Maxwell-like function, the low-temperature component being associated with ponderomotive electrons and the high-temperature component with electrons accelerated inside the plasma channel via the DLA mechanism.

To analyse the acceleration regime realized in the first stage, a set of test electrons was introduced. These particles were distributed over a broad range of initial longitudinal and transverse momenta and were assigned a small statistical weight to preclude any perturbation of the plasma dynamics. The injection coordinates were \(x_0 = (25-27)\lambda\), \(r_0 = 0.5\lambda\), and the injection time was \(t_0 = 200\lambda/c\), these parameters being selected on the basis of the acceleration dynamics observed in the simulation so as to replicate the starting conditions of the electrons that are naturally accelerated in the first stage (see Fig.~\ref{Figure1}b,c). Fig.~\ref{Figure1}e displays the maximum energy of these test electrons as a function of their initial momenta. The colour map demonstrates that efficient acceleration in the region of low initial energies, which indicates that the chosen parameters are well suited for the implementation of the first stage. The acceleration pattern observed in this figure is compared with the theoretical DLA model later in Sec.~\ref{Sec3}.

\section{Second-stage acceleration}

The second stage was simulated with a larger computational domain \(490\lambda \times 100\lambda\) and a total simulation time \(650\lambda/c\), required by the transverse and longitudinal size of the injected electron bunch and by the longer propagation distance over which the pre-accelerated electrons gain energy. The plasma was hydrogen with a lower density, \(n_{e} = 0.01n_{cr}\), and the profile was shifted along the longitudinal coordinate with the density up-ramp located at \(x = 171\lambda\), as required by the large longitudinal extent of the injected electron bunch, which is approximately \(150\lambda\). The use of a low-density hydrogen stage following a high-density injection stage is a well-established configuration in LWFA, where a high-Z gas at higher density provides efficient injection, while a low-Z gas at lower density extends the acceleration length and preserves the beam quality \cite{tomkus2024laser, gardner2025implementing, maity2025coupling}. We, therefore, adopted the same approach here. The laser pulse parameters were the same as in the first stage (\(a_0 = 6\), \(\tau = 200\) fs), except that the waist was increased to \(w = 8\lambda\). This modification was motivated by our previous theoretical analysis of DLA, which indicates that efficient acceleration in the second stage requires a lower phase velocity in the plasma channel \cite{starodubtseva2023cascade}; a larger waist enhances the channel depletion, reduces the on-axis plasma density, and thereby increases the phase velocity of the accelerating structure.

\begin{figure*}
\centering
\includegraphics[width=1.0\textwidth]{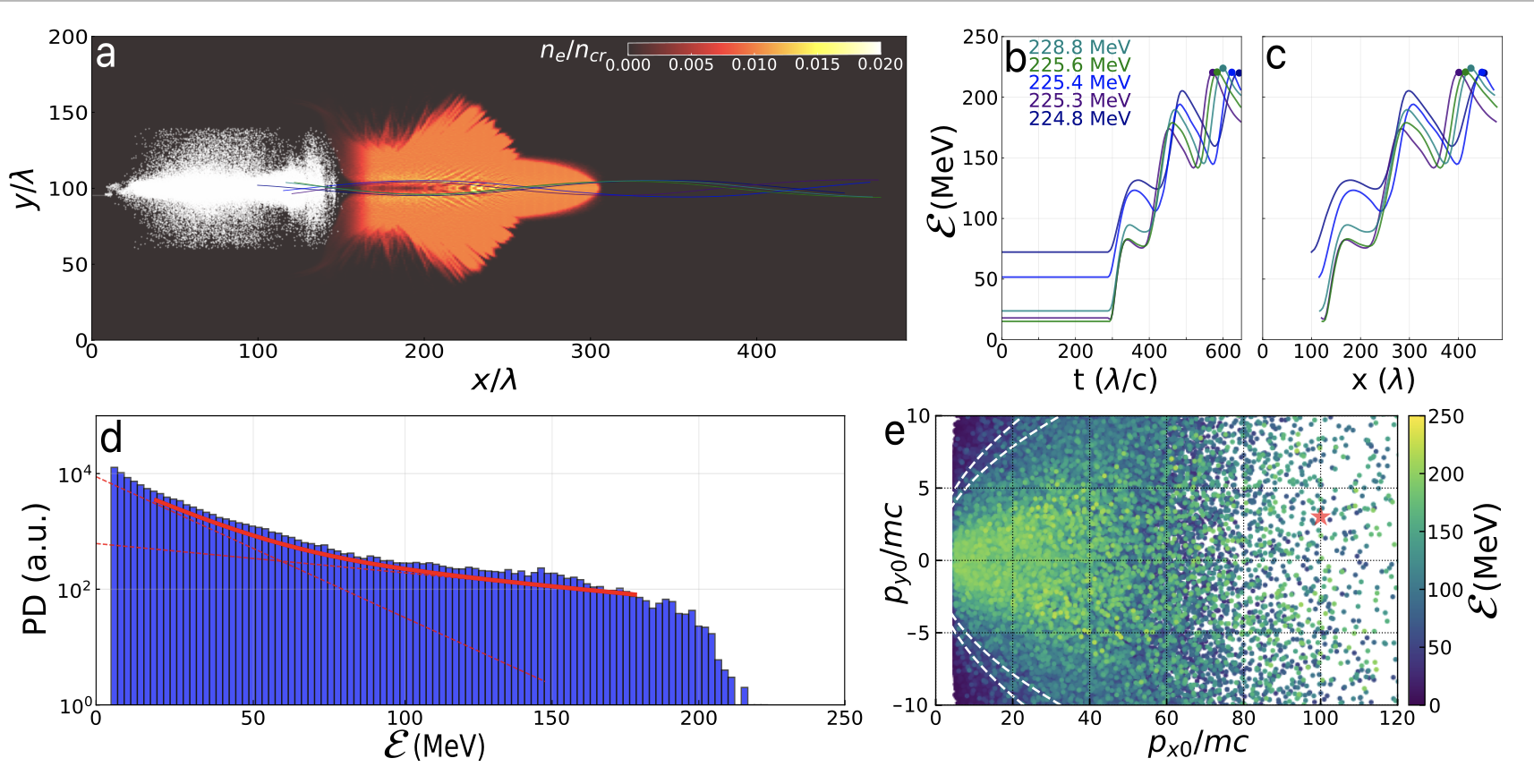}
\caption{Second-stage acceleration results. (a) Plasma density snapshot at the instant, overlaid with trajectories of the five most energetic electrons. The white scatters show the injected electron bunch with \(p_x/mc > 5\) at the initial instant. (b) Energy evolution over time for the five most energetic electrons; the maximum energy reached by each electron is indicated in the legend, dots indicate the instants of maximum energy. (c) Energy of the same electrons as a function of longitudinal position; dots indicate the positions of maximum energy. (d) Energy spectrum of accelerated electrons above 5 MeV (blue histogram) and fit \(dN/d\mathcal{E} \propto A_1 \exp(-\mathcal{E}/T_1) + A_2 \exp(-\mathcal{E}/T_2)\) (solid red curve), yielding \(T_1 = 18.4\) MeV and \(T_2 = 88.3\) MeV (dashed red curves). (e) Maximum energy of electrons as a function of initial transverse and longitudinal momenta. The white dashed curves mark the analytical boundaries of the different electron regime. The red star indicates the point \((p_x/mc, p_y/mc) = (100, 3)\) selected for phase portrait analysis (see Sec.~\ref{Sec3}). Simulation domain: \(490\lambda \times 100\lambda\), total time \(650\lambda/c\).}
\label{Figure2}
\end{figure*}

The results of the second-stage acceleration are presented in Fig.~\ref{Figure2}. Fig.~\ref{Figure2}a shows a snapshot at $t=360\lambda/c$, overlaid with the trajectories of the five most energetic electrons. The white scatters show the injected electron bunch with \(p_x/mc > 5\) at the initial instant. The trajectories continue to exhibit betatron oscillations, confirming that the DLA mechanism remains operative for the pre-accelerated electrons. Fig.~\ref{Figure2}b illustrates the temporal evolution of the energy for the five most energetic electrons, demonstrating the acceleration process over \(300\lambda/c\); the maximum energy reached by each electron is indicated in the legend. Fig.~\ref{Figure2}c shows the energy of the same electrons as a function of their longitudinal coordinate; acceleration is observed to occur over \(\sim 300\lambda\). The trajectories and energy evolution confirm that the electrons undergo acceleration up to \(\sim 225\) MeV. Fig.~\ref{Figure2}d displays the energy spectrum of the electron bunch after the second stage (blue histogram, \(\mathcal{E} > 5\) MeV). The spectrum is well described by a two-temperature distribution, with effective temperatures of \(T_1 = 18.4\) MeV and \(T_2 = 88.3\) MeV. The electron beam charge after the second stage was \(4.55\) nC (\(\mathcal{E} > 5\) MeV), which also demonstrates efficient cascaded acceleration. The self-injected charge and electron energies in the second stage were also analyzed. The maximum energy reached 123 MeV, and the charge of such electrons with energies above 5 MeV was 1.14 nC, which demonstrates the high efficiency of multi-stage DLA acceleration  compared with single-stage one.

Fig.~\ref{Figure2}e presents the maximum energy of electrons as a function of their transverse and longitudinal initial momenta. The colour map demonstrates efficient acceleration of electrons with high initial energies, which indicates that the chosen parameters are well suited for the implementation of the second stage. At the same time, one can also observe electrons that do not gain significant energy. This can be attributed to two main reasons: either the electrons have a large transverse momentum and consequently escape far beyond the channel boundaries, where they cannot be captured by the accelerating structure, or they are injected into a decelerating phase, which prevents efficient energy gain. The acceleration pattern observed in the Fig. ~\ref{Figure2}e is compared with the theoretical DLA model later in Sec.~\ref{Sec3}.

\section{Comparison with the DLA Model}
\label{Sec3}

The simplified 2D model of DLA describes the electron motion in the combined fields of the laser pulse and the quasistatic plasma channel. The model developed in our previous work~\cite{starodubtseva2023low} is suitable for small oscillation amplitudes and for ionization injection in DLA. It cannot, however, be applied directly to the present problem, because it assumes a plasma channel that is infinite in the transverse direction and considers electrons starting strictly on the channel axis. In the multi-stage DLA scheme considered here, and especially in the second stage, the injected electron bunch is broad in the transverse direction and the betatron oscillation amplitudes are large, so that both assumptions are violated. Two modifications are therefore required: (i) the transverse attenuation of the channel fields must be taken into account, and (ii) off-axis injection must be considered.

The channel fields arise from charge separation caused by the laser ponderomotive force, which expels electrons from the axis and leaves an ion column that generates a radial electric field, while the longitudinal electron current produces an azimuthal magnetic field. Since the plasma remains globally neutral, the charge and current densities must integrate to zero over the transverse cross-section. A simple analytical profile satisfying this condition is a Gaussian distribution modulated by a quadratic term:
\begin{equation}
\rho(r),\, j_x(r) \propto e^{-r^2/r_0^2}\left(1 - \frac{r^2}{r_0^2}\right),
\label{eq:density}
\end{equation}
where \(r_0\) is the characteristic channel radius, and \(y\) is the transverse coordinate. This form reflects the overall neutrality of the plasma and the compensation of the fast-electron current by the return plasma current \cite{tsakiris2000laser}.

Applying Gauss's law and Amp\`ere's law to the charge and current distribution~\eqref{eq:density}, and assuming that the electron motion is confined to the plane \(z = 0\), yields the static channel fields:
\begin{equation}
\begin{aligned}
E_y^s &= -2\pi e (n_e - n_{e_{\mathrm{in}}})\, y\, e^{-y^2/r_0^2}, \\
B_z^s &= -\frac{2\pi e n_{e_{\mathrm{in}}} v_x}{c}\, y\, e^{-y^2/r_0^2},
\end{aligned}
\label{eq:fields}
\end{equation}
where \(n_{e_{\mathrm{in}}}\) is the electron density inside the channel, \(e\) is the electron charge, \(v_x\) is the longitudinal electron velocity, and \(c\) is the speed of light. For \(y \ll r_0\), Eqs.~\eqref{eq:fields} are reduced to the linear fields of a homogeneous channel used in Ref.~\cite{starodubtseva2023low}, whereas for \(y \sim r_0\) they describe a channel of finite transverse size.

The laser pulse is modeled as a uniform plane wave:
\begin{equation}
\begin{aligned}
E_y^L &= \frac{a_0 m c \omega}{e}\cos(\omega t - kx), \\
B_z^L &= \frac{c}{v_{\mathrm{ph}}}\,\frac{a_0 m c \omega}{e}\cos(\omega t - kx),
\end{aligned}
\label{eq:laser}
\end{equation}
where \(\omega\) and \(k\) are the laser frequency and wavenumber, \(x\) is the longitudinal coordinate, \(v_{\mathrm{ph}}\) is the phase velocity of the laser pulse in the plasma channel, and \(m\) is the electron mass.

The electron motion in this field configuration is described by the relativistic equations of motion for the longitudinal and transverse momenta $p_x$ and $p_y$:
\begin{equation}
\begin{cases}
\dfrac{dp_x}{dt} = -e\left(E_y^s + E_y^L\right) - \dfrac{e v_x}{c}\left(B_z^s + B_z^L\right), \\[10pt]
\dfrac{dp_y}{dt} = -\dfrac{e v_y}{c}\left(B_z^s + B_z^L\right),
\end{cases}
\label{eq:motion}
\end{equation}
where \(v_y\) is the transverse electron velocity.

The Gaussian form of the channel fields (see Eq.~\eqref{eq:fields}) changes the integral of motion with respect to the previous model ~\cite{starodubtseva2023low}:
\begin{equation}
I_0 = \gamma - \frac{p_x}{\eta m c}
+ \mathcal{C}\left[\exp\left(-\frac{y^2}{r_0^2}\right) - 1\right] = \mathrm{const},
\label{eq:integral}
\end{equation}
where \(\eta = c/v_{\mathrm{ph}}\) is the refractive index inside the plasma channel and
\begin{equation}
\mathcal{C} = \frac{\pi e^2 r_0^2}{m c^2}\left[(n_{e_{\mathrm{in}}} - n_e) + \frac{n_{e_{\mathrm{in}}}}{\eta}\right].
\label{eq:C}
\end{equation}
On the axis and in its vicinity (\(y \ll r_0\)), the exponential term in Eq.~\eqref{eq:integral} can be expanded, and the integral reduces to the form obtained previously in Ref.~\cite{starodubtseva2023low}. Off axis, however, the additional term becomes significant and accounts for the finite transverse size of the channel.

Under the WKB approximation and averaging over the fast oscillations of the electron energy, the system admits an analytical solution that relates the electron energy \(\mathcal{E}\) to the relative phase \(\Phi\). The phase trajectories are given by
\begin{equation}
\sin\Phi = F(\mathcal{E}) - \alpha,
\label{eq:solution}
\end{equation}
where \(\Phi = \omega t - \int \omega_b\, dt - kx + \Phi_0\) is the relative phase between the laser field and the betatron oscillations, with \(\Phi_0\) being the initial phase, and the function \(F(\mathcal{E})\) is determined by the plasma and laser parameters and by the integral of motion Eq. ~\eqref{eq:integral}, and \(\alpha = F(\mathcal{E}_0) - \sin\Phi_0\) is fixed by the initial conditions (\(\mathcal{E}_0\) is the initial electron energy). The function \(F(\mathcal{E})\) and the expressions for the critical points retain the same form as in Ref.~\cite{starodubtseva2023low}, up to the replacement of the integral of motion by the new expression Eq.~\eqref{eq:integral}, which accounts for the Gaussian channel fields Eq.~\eqref{eq:fields} and for off-axis electron starting \(y_0\). The critical values of the integral of motion that determine the topology of the phase portraits, \(I_{0cr}\) and \(I_{0f}\), are governed only by the plasma and laser parameters and therefore retain the same form as in Ref.~\cite{starodubtseva2023low}.

\begin{figure}
\centering
\includegraphics[width=0.5\textwidth]{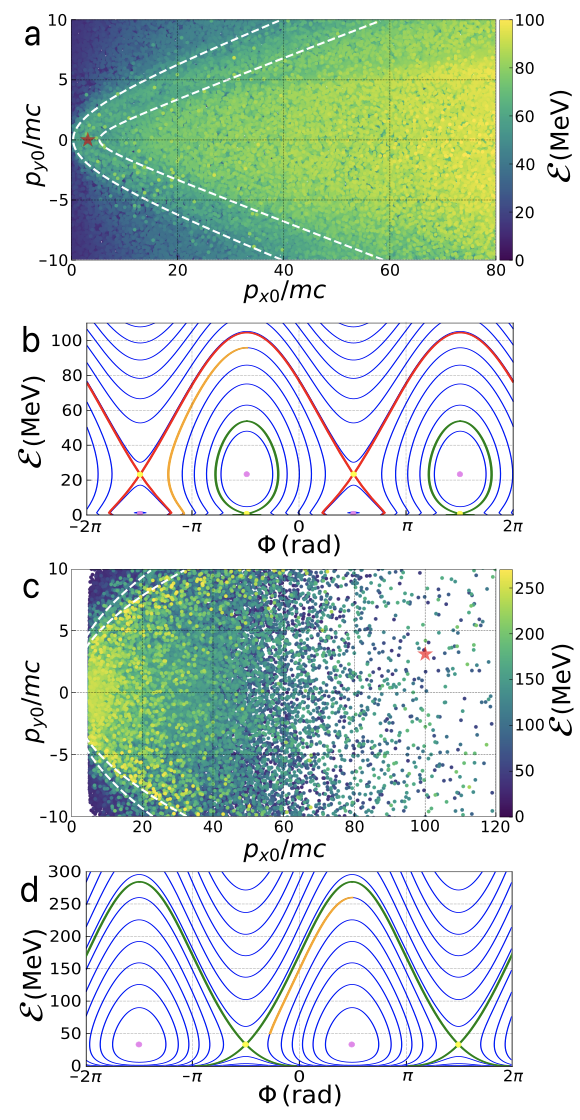}
\caption{Results of the DLA analytical model using initial parameters and plasma channel radii extracted from the PIC-simulations. (a) Maximum energy as a function of initial momenta for the first stage, obtained at \(v_{ph} = 1.0161c\) and \(n_e = 0.056n_{cr}\) over \(150\lambda/c\). The white dashed curves mark the analytical boundaries of the different acceleration regime, and the red star indicates the point \((p_x/mc, p_y/mc) = (3, 0)\) selected for the phase portrait analysis. (b) Analytical phase portrait in the \((\mathcal{E},\Phi)\) plane for the point \((p_x/mc, p_y/mc) = (3, 0)\), revealing the efficient acceleration of low-energy electrons. The orange curve represents analytical phase trajectory within this regime, illustrating initially low-energy electron energy gain. (c) Same as (a) but for the second stage, at \(v_{ph} = 1.0016c\) and \(n_e = 0.01n_{cr}\) over \(300\lambda/c\); the red star indicates the point \((p_x/mc, p_y/mc) = (100, 3)\) selected for the phase portrait analysis. (d) Analytical phase portrait for the point \((p_x/mc, p_y/mc) = (100, 3)\), showing the regime of efficient acceleration of pre-accelerated electrons. The orange curve represents analytical phase trajectory within this regime, illustrating initially pre-accelerated electron energy gain.}
\label{Figure3}
\end{figure}

To validate the modified model against the PIC-simulations, the full relativistic equations of motion Eq.~\eqref{eq:motion} with the Gaussian channel fields Eq.~\eqref{eq:fields} were integrated numerically. For the first stage, the numerical integration was performed at the plasma density \(n_e = 0.056 n_{cr}\) and the phase velocity \(v_{ph} = 1.0161c\), measured from the PIC simulation using the technique proposed in our recent work~\cite{starodubtseva2026phase}, over a time interval of \(150\lambda/c\), corresponding to the acceleration duration observed in Fig.~\ref{Figure1}b. Based on the PIC-simulations, the characteristic channel radius was set to \(r_0 = 3.0\lambda\). To accurately reproduce the PIC-simulation results, the initial states of the test electrons -- both longitudinal and transverse coordinates as well as their momenta -- were taken directly from the PIC-simulation. The resulting maximum energy as a function of initial momenta is shown in Fig.~\ref{Figure3}a, which can be directly compared with the PIC result in Fig.~\ref{Figure1}e. The white dashed curves mark the analytical boundaries of the low-energy acceleration regime, which are determined by the critical values of the integral of motion, \(I_{0cr}\) and \(I_{0f}\). These critical values separate the phase space regions with different topologies: \(I_{0cr}\) corresponds to the separatrix reconnection, and \(I_{0f}\) to the merging and disappearance of the singular points. The corresponding expressions are given in Ref.~\cite{starodubtseva2023low}.

For the specific phase portrait analysis of the point \((p_x/mc, p_y/mc) = (3, 0)\), marked by a star in Fig.~\ref{Figure1}e and in Fig.~\ref{Figure3}a, the initial transverse coordinate was set to \(y_0 = 0.5\lambda\), matching the off-axis injection of the test electrons in the PIC-simulation. The corresponding phase portrait is presented in Fig.~\ref{Figure3}b. The orange trajectory illustrates the capture of a low-energy electron into the accelerating phase and its energy gain up to large values.

For the second stage, the numerical integration was performed at \(n_e = 0.01 n_{cr}\) and \(v_{ph} = 1.0016c\) over a time interval of \(300\lambda/c\). Based on the PIC simulations, the characteristic channel radius was increased to \(r_0 = 6.0\lambda\). Similar to the first stage, the initial states of the test electrons were extracted directly from the PIC simulations. This accounts for the broad transverse profile of the injected beam, with transverse coordinates spanning a much wider range compared to the first stage. The resulting maximum energy map is shown in Fig.~\ref{Figure3}b.  The white dashed curves mark the analytical boundaries of the different electron acceleration regime. The phase portrait for the point \((p_x/mc, p_y/mc) = (100, 3)\), marked by a red star in Fig.~\ref{Figure2}e and in Fig.~\ref{Figure3}c, is presented in Fig.~\ref{Figure3}d. For this point, the initial transverse coordinate was set to \(y_0 = 3.5\lambda\), matching the off-axis injection of the injected electrons in the second stage.The orange trajectory illustrates the capture and continuous acceleration of a pre-accelerated electron.

The comparison between the analytical model and the PIC-simulations demonstrates that the modified DLA model correctly describes the observed acceleration dynamics in both stages. The regions of efficient acceleration in the maximum energy maps coincide with the predictions of the model, and the phase portraits reproduce the characteristic features of the energy gain for low-energy and pre-accelerated electrons. 

\section{Conclusion}

In this work, we presented the first numerical demonstration of two-stage direct laser acceleration (DLA) of electrons in quasi-3D particle-in-cell (PIC) simulations in cylindrical geometry with azimuthal mode decomposition. The first stage was designed to accelerate low-energy electrons (\(a_0 = 6\), \(\tau = 200~\mathrm{fs}\), \(w = 4\lambda\), \(n_e = 0.056 n_{cr}\), \(\lambda = 0.8 \mu\mathrm{m}\), with nitrogen as the target gas), achieving a maximum energy of approximately 94~MeV with a charge of 6.04~nC (\(\mathcal{E} > 5\) MeV) and a two-temperature energy spectrum with \(T_1 = 7.0\) MeV and \(T_2 = 16.0\) MeV. The resulting electron bunch was then injected into the second stage, where the laser waist was increased to \(w = 8\lambda\) and hydrogen was used at a lower density \(n_e = 0.01 n_{cr}\) to efficiently accelerate the pre-accelerated electrons. After the second stage, the electrons reached energies up to 225~MeV, with a two-temperature spectrum (\(T_1 = 18.4\) MeV and \(T_2 = 88.3\) MeV) and a preserved charge of 4.55~nC (\(\mathcal{E} > 5\) MeV). In contrast, the self-injected electrons in the second stage only reached a maximum energy of 123~MeV with a charge of 1.14~nC (\(\mathcal{E} > 5\) MeV). The simulation results confirm the feasibility of sequential DLA acceleration and demonstrate that the DLA mechanism can be successfully implemented in a multi-stage configuration. The charge and maximum energies of the injected electron beam after the second stage are significantly higher than those of the self-injected electrons, highlighting the key advantage of the cascaded scheme.

To provide a theoretical framework for the observed acceleration dynamics, we employed a modified analytical DLA model. Building upon our previous work~\cite{starodubtseva2023low}, this upgraded model accounts for the finite transverse size of the electron beam by incorporating a Gaussian attenuation of the plasma channel fields and considering electron injection away from the axis of the plasma channel.

To validate the analytical predictions, the full relativistic equations of motion were integrated numerically using the initial parameters and plasma channel radii (\(r_0 = 3.0\lambda\) and \(r_0 = 6.0\lambda\) for the first and second stages, respectively) extracted directly from the PIC-simulations. The resulting maximum energy maps demonstrate good agreement with the PIC-simulations results, successfully reproducing both the qualitative features and the quantitative energy values. 

However, the analytical phase portrait for a representative electron from the first stage demonstrates that the actual acceleration length in the PIC simulation is insufficient to reach the maximum energy theoretically achievable with these given laser and plasma parameters. While the PIC electron reaches approximately 90 MeV, the phase portrait indicates a theoretical limit of up to \(\sim 200\) MeV. This discrepancy arises because the plasma channel dissipates rapidly at the high density of the first stage (\(n_e = 0.056 n_{cr}\)), prematurely terminating the acceleration before the electron can fully extract the available energy. We anticipate that if the channel lifetime could be extended—for instance, by pre-forming the channel using a capillary discharge—the high electron energies predicted by the phase portrait could indeed be achieved with the same laser and plasma parameters.

The obtained results are of significant importance for the development of multi-stage accelerator schemes. The PIC simulations have shown that DLA is particularly well suited for cascaded acceleration: it provides not only efficient energy gain but also a substantially higher charge compared to laser wakefield acceleration (LWFA). This high charge, combined with the ability to independently optimize each stage, makes DLA a promising candidate for compact, high-current electron sources. Furthermore, the modified analytical model allows for a more realistic estimation of the optimal laser and plasma parameters for each stage, taking into account finite channel widths and beam sizes. This predictive capability significantly simplifies the design of multi-stage DLA schemes, enabling researchers to quickly assess the feasibility of different configurations before running expensive PIC-simulations. Overall, this work establishes a solid foundation for the future experimental implementation of multi-stage DLA accelerators.

\begin{acknowledgments}
This work was supported by Russian Science Foundation grant №22-79-10087P with the use of equipment purchased under support of National project Science and Universities of Ministry of Science and Higher Education of the Russian Federation. E.M. acknowledges Foundation for the advancement of theoretical physics “BASIS” for the financial support.
\end{acknowledgments}

\section*{Data Availability Statement}
The data that support the findings of this study are available from the corresponding author upon reasonable request.

\bibliography{main}

\end{document}